\documentclass[aps,pre,reprint,amsmath,amssymb,floatfix]{revtex4-2}
\usepackage{graphicx}
\usepackage{xcolor}
\usepackage{overpic}
\usepackage{booktabs}
\usepackage{makecell}
\graphicspath{{figures/}}

\begin{document}

\title{Random close packing at extreme size ratios with an Adam-based inflation protocol}
\author{K.W.~Desmond}
\email{kdesmond.physics@gmail.com}
\affiliation{Independent Researcher}
\date{\today}

\begin{abstract}
We present \texttt{rcpgenerator}, an openly available code for generating $d$-dimensional dense, disordered, non-overlapping close packings from an arbitrary prescribed list of particle diameters. The method adapts the Clarke--Wiley inflation protocol, but instead uses the Adam optimizer to relax the particle configuration. Typically, particle coordinates are advanced with a single, global step size, which must shrink as the size ratio $S\equiv D_{\max}/D_{\min}$ grows, generally stalling the optimization. Adam instead gives each coordinate its own adaptive step size, stabilizing the optimization time across a broader range of $S$. We demonstrate this in three-dimensional periodic tests that reach $S\sim5\times10^{5}$ for a continuous lognormal distribution ($N\sim10^{6}$ diameters) and particle numbers up to $N\approx5.6\times10^{6}$ for power-law distributions, with the densest packings reaching $\phi\simeq0.87$, each completed in minutes to hours on a multicore machine. Across truncated-lognormal, truncated-power-law, and Weibull distributions, the resulting $\phi$ reproduces trends such as the locations of peaks and knees with distribution shape and $S$ found in prior numerical results and in the parameter-free Farr--Groot prediction, with a remaining offset typically $0.005$--$0.01$. Additionally, results are commensurate with multimodal packing densities measured in vibrated-bed experiments. The code and the complete per-case census behind every figure are released with the paper.	
\end{abstract}

\maketitle

\section{Introduction}\label{sec:intro}

Real particulate systems are almost never monodisperse---granular media, powders, sediments, and colloids carry broad size distributions spanning size ratios from tens to thousands~\cite{voivret_2009,folk_ward_1957,pusey_1987}. Their structure is central to a range of applications, including additive manufacturing~\cite{bai_2017,young_2022}, particle-resolved fluid and granular simulation~\cite{jerier_2010,novikov_besseron_2025}, and flow through porous media~\cite{garcia_2009}, and in each, the packing density largely governs the mechanical and transport properties. Broad polydispersity is therefore physically common, but remains among the hardest to generate numerically.

Many applications only require the final packing structure, not the trajectory to it, and may be produced through dynamic or optimization-based solvers, such as molecular dynamics or energy minimization. However, as the ratio of largest to smallest diameter increases, resolving the increasingly disparate length and stiffness scales, spanning several orders of magnitude for the broadest distributions, grows rapidly more costly~\cite{lubachevsky_stillinger_1990,ogarko_luding_2012,shire_hanley_stratford_2021,monti_2022}, and most numerical work has remained at size ratios $S\equiv D_{\max}/D_{\min}$ of order ten to a hundred~\cite{shimamoto_2023,oquendo_estrada_2022,bolton_lum_2026, anzivino_2023}. Reaching much broader distributions then requires a solver better suited to the ill-conditioning they create.

Our approach builds on the Clarke--Wiley protocol~\cite{clarke_wiley_1987}, as adapted in numerous studies~\cite{desmond_weeks_2009, desmond_weeks_2014,xu_blawzdziewicz_ohern_2005}, which generates dense, disordered packings by inflating particles from a dilute configuration while relaxing the resulting overlaps. At large $S$, the minimization becomes increasingly ill-conditioned because the contact network contains widely separated force and curvature scales, analogous to the anisotropic objectives encountered in neural-network optimization~\cite{sagun_evci_2017, pennington_bahri_2017,ghorbani_krishnan_xiao_2019}. Adaptive per-coordinate optimizers were developed for such objectives; here we use Adam~\cite{kingma_ba_2015} for the joint relaxation of particle positions and a common diameter scale. The update to each coordinate is rescaled using recent gradient statistics, without Hessian information or line searches, keeping the per-iteration computation cost low. We implement this Adam-based inflation protocol in the openly available code, \texttt{rcpgenerator}, that accepts a prescribed list of $N$ relative diameters, a spatial dimension, and a boundary geometry. Since coordinates adapt independently, the number of updates needed to reach the stopping criterion depends only weakly on $S$; for instance, power-law systems up to $S=200$ and lognormal systems above $S=10^{4}$ with $N \sim 10^{6}$ require order $10^{5}$ updates. Throughout, we use rcp to denote the dense, disordered, non-overlapping state produced by this protocol, which is neither strictly nor collectively jammed~\cite{torquato_truskett_debenedetti_2000,torquato_stillinger_2010}.

While Adam can reach dense packings efficiently, we test whether the packing fractions $\phi$ it produces are commensurate with other findings where results exist at modest size ratios (numerical~\cite{anzivino_2023,monti_2022} and parameter-free Farr--Groot theory~\cite{farr_groot_2009,farr_2013}). We compare across a broad range of system size and polydispersity, for three distribution families: truncated lognormal, truncated power law, and Weibull. The cases tested here reach $S\sim5\times10^{5}$ for the lognormal family (up to $N\approx3.1\times10^{6}$) and $N\approx5.6\times10^{6}$ particles for the power law (at $S=200$) in hours. Across the families, $\phi$ reproduces the predicted dependence on $P(D)$ shape and $S$, including the same peaks, knees, and trends, with small systematic offsets: typically $0.005$--$0.01$ below the references, increasing to about $0.03$ for the densest power-law packings ($\phi\simeq0.87$), while the Weibull results lie at or slightly above the Farr--Groot prediction. Additionally, in Brouwers~\cite{brouwers_2025}, binary packings generated by \texttt{rcpgenerator} showed broad agreement with theory and prior simulations.

\begin{figure*}[t]\centering
	\includegraphics[width=0.8\textwidth]{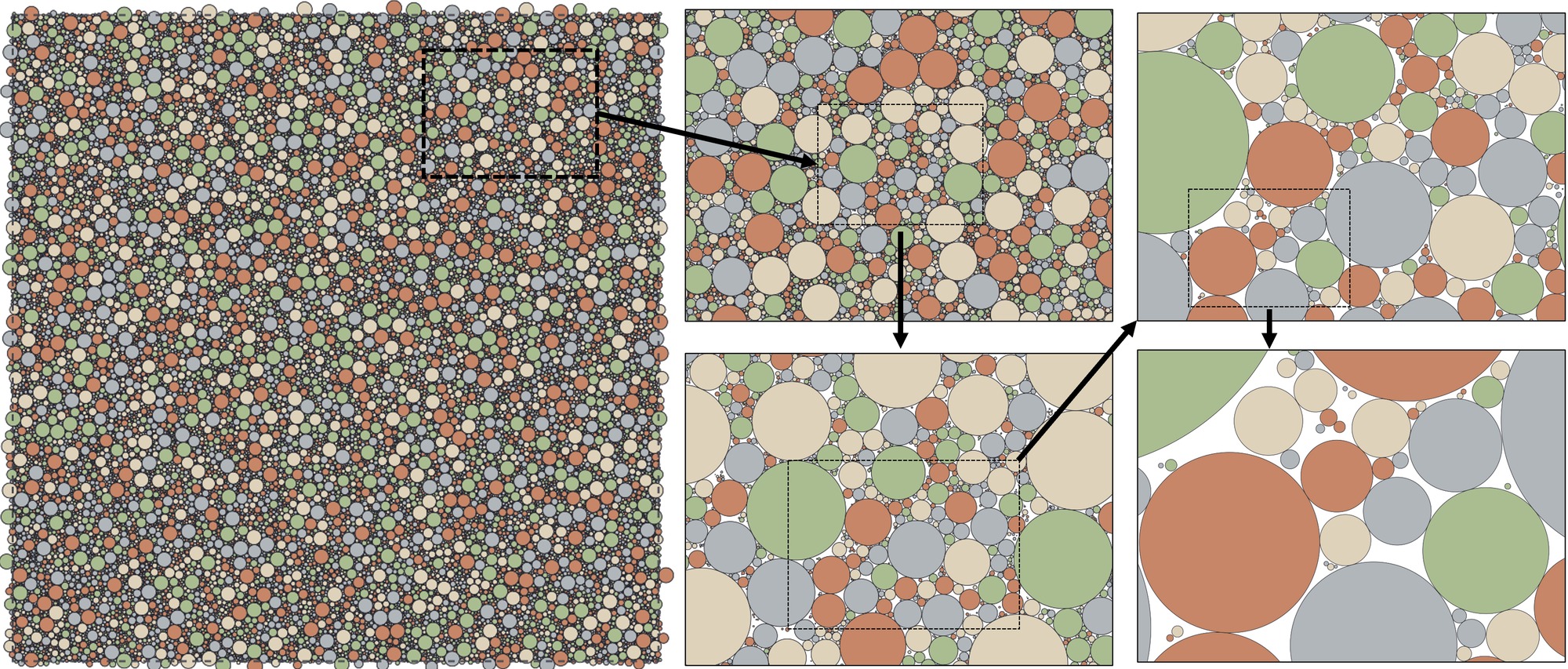}
	\caption{2D rcp of a truncated-lognormal distribution ($\alpha=1.3$), with $N=10^{5}$, $S\approx8.4\times10^{3}$, and $\phi\approx0.91$. The full periodic cell and four successive close-up panels of the indicated regions highlight roughly four decades in length scale, resolving the smallest particles only in the final panel.}
	\label{fig:extreme}
\end{figure*}

\section{Method and capabilities}\label{sec:algo}

\texttt{rcpgenerator} implements the Clarke--Wiley inflation protocol as adapted by Desmond and Weeks~\cite{desmond_weeks_2009,desmond_weeks_2014}. The solver begins from a dilute, non-overlapping configuration of $N$ particles with a prescribed list of positive relative diameters $\{D_i^0\}$. Their ratios remain fixed throughout the optimization protocol, while a single global scale $\kappa$ sets the instantaneous diameters, $D_i=\kappa D_i^0$. The particle positions and $\kappa$ are optimized jointly under a finite-range soft repulsion and a pressure-like term that favors growth. The pressure is reduced through a fixed annealing schedule, and the final relaxation terminates when the maximum fractional pair overlap falls below a prescribed tolerance. Further details of the energy functional, Adam update, annealing schedule, and stopping criteria are given in Appendix~\ref{app:impl}.

The code accepts any finite diameter list and imposes no parametric form on it. For the continuous distributions studied here, the diameters are chosen as equal-probability inverse-CDF quantiles at cumulative probabilities $(i+1/2)/N$. The overall scale must be compatible with the selected domain; in a periodic cell, in particular, the largest particle must fit without interacting with its own periodic image. The joint minimization is performed with the Adam optimizer~\cite{kingma_ba_2015}, whose per-coordinate adaptive steps accommodate the wide range of force and curvature scales that broad polydispersity produces; multiscale neighbor management and the force-evaluation inner loop are described in Appendix~\ref{app:impl}. Figure~\ref{fig:extreme} shows a two-dimensional packing viewed through successive magnifications spanning roughly four decades in length scale. Each panel shows newly resolved particles filling structure invisible at the coarser scale and never overlapping.

\begin{figure}[b]\centering
	\begin{minipage}[c]{0.49\linewidth}\includegraphics[width=\linewidth]{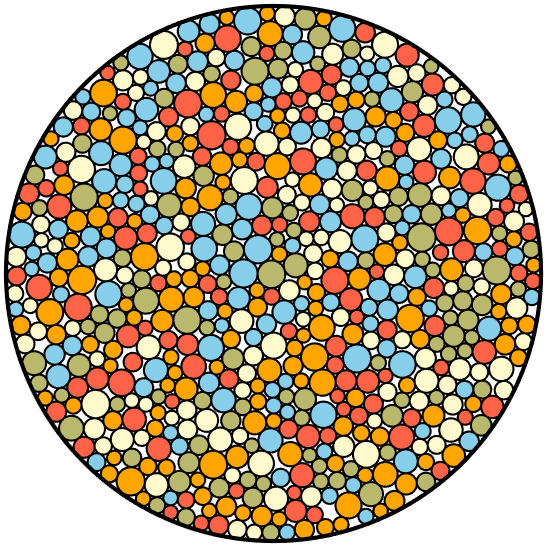}\end{minipage}\hfill
	\begin{minipage}[c]{0.49\linewidth}\includegraphics[width=\linewidth]{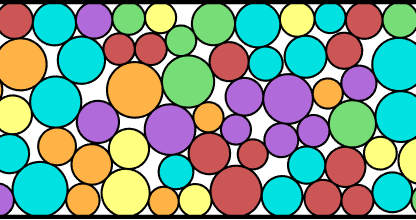}\end{minipage}\\[2pt]
	\begin{minipage}[c]{0.49\linewidth}\includegraphics[width=\linewidth]{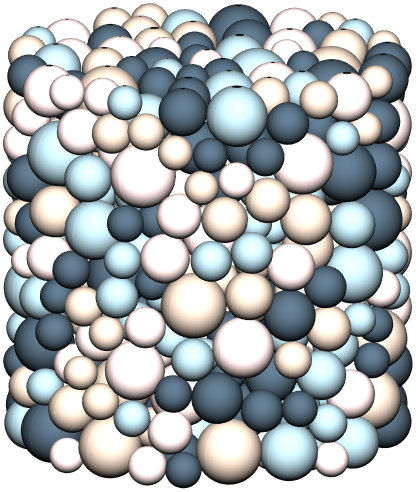}\end{minipage}\hfill
	\begin{minipage}[c]{0.49\linewidth}\includegraphics[width=\linewidth]{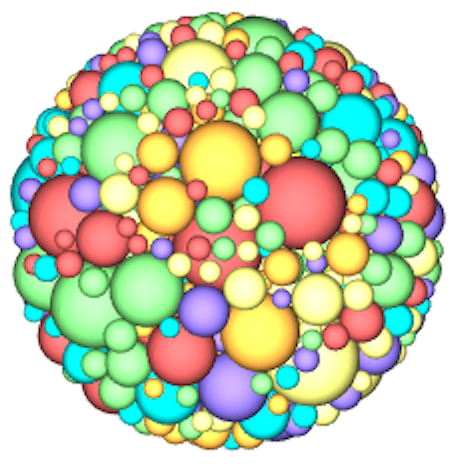}\end{minipage}
	\caption{Examples of curved and hard boundary conditions beyond periodic rectilinear supported by the code.}
	\label{fig:capabilities}
\end{figure}

Each coordinate direction may be periodic or bounded by a flat or curved hard wall, allowing rectilinear, circular, cylindrical, and spherical domains within the same formulation (Fig.~\ref{fig:capabilities}). The spatial dimension $d$ is also a free input, although the quantitative comparisons reported here are restricted to three-dimensional periodic packings.

At termination the configuration retains a small residual overlap, which we remove by uniformly reducing $\kappa$ until the last overlap vanishes, enforcing strict non-overlap at the cost of lowering $\phi$ by $\approx9\times10^{-4}$ on average. The resulting dense, disordered, non-overlapping configuration is what we call the random-close-packed state, and its solid-volume fraction is the shrink-to-contact packing fraction $\phi_{\rm rcp}(P(D),N)$, written $\phi$ throughout.

Table~\ref{tab:cap} lists representative packings, the largest at $N\approx1.05\times10^{6}$ and $S\approx2.3\times10^{5}$, generated in minutes to hours; the full census reaches $N\approx5.6\times10^{6}$ and $S\approx5\times10^{5}$. Additionally, in a controlled single-thread benchmark spanning 
$N = 5\times10^{3}$ to $N = 1.6\times10^{5}$, the measured generation time scales as
$t(N)\propto N^{1.34}$ (released in \texttt{scaling\_benchmark.csv}), corresponding to an approximately $2.5$-fold increase
in compute time when $N$ is doubled. We run no controlled comparison among optimizers; the timings characterize the complete Adam-based implementation. The scales here were set by the scope and computational resources of this study.

\begin{table}[t]\centering
	\begin{tabular}{rrrrr}
\toprule
$\alpha$ & $S$ & $N$ & $\phi$ & \shortstack{compute\\ time (min)} \\
\midrule
0.10 & $3$ & $2{,}000$ & 0.641 & 0.0 \\
0.50 & $116$ & $2{,}000$ & 0.698 & 0.1 \\
0.70 & $773$ & $8{,}784$ & 0.735 & 0.2 \\
0.90 & $5{,}167$ & $75{,}966$ & 0.772 & 2.9 \\
1.00 & $13{,}360$ & $195{,}929$ & 0.788 & 12.8 \\
1.10 & $34{,}544$ & $413{,}799$ & 0.799 & 35.9 \\
1.25 & $143{,}631$ & $883{,}556$ & 0.809 & 98.3 \\
1.30 & $230{,}960$ & $1{,}049{,}699$ & 0.813 & 124.2 \\
\bottomrule
\end{tabular}

	\caption{Representative packings showing the size ratio $S\equiv D_{\max}/D_{\min}$, particle number $N$, packing fraction $\phi$, and compute time. Times are for one packing using 14 CPU cores and are included only to indicate computational scale.}
	\label{tab:cap}
\end{table}

\section{Comparison of continuous $P(D)$ with theory and simulation}\label{sec:results}
We run the generator across three continuous distribution families: truncated lognormal, truncated power law, and Weibull. Each is chosen to span a wide range of distribution shape, size ratio, and packing fraction, and we compare against molecular-dynamics simulations where available. The effort comprises $1166$ independent random close packings and more than $3.5\times10^{8}$ packed spheres in aggregate, $128$ of them above one million particles (the largest near $N\approx5.6\times10^{6}$) and all produced over a few days on a $44$-core cloud virtual machine, an indication of the throughput the generator sustains at these sizes. Table~\ref{tab:census} provides a list of the parameter range.

\begin{table}[t]\centering
	\begin{tabular}{lccccc}
		\toprule
		Family & Cases & $N_{\min}$ & $N_{\max}$ & $S_{\min}$ & $S_{\max}$ \\
		\midrule
		Lognormal & 63  & $2.0\times10^{3}$ & $3.1\times10^{6}$ & 2   & $5.0\times10^{5}$ \\
		Power law & 255 & $2.0\times10^{3}$ & $5.6\times10^{6}$ & 5   & 200 \\
		Weibull              & 10  & $2.6\times10^{3}$ & $3.1\times10^{6}$ & 300 & 300 \\
		\bottomrule
	\end{tabular}
	\caption{Parameter range spanned by each continuous family. \emph{Cases} is the number of distinct $P(D)$; each packed at several system sizes and seeds.}
	\label{tab:census}
\end{table}

\subsection{Lognormal distribution}\label{sec:val}
\begin{figure*}[tb]\centering
	\includegraphics[width=0.8\textwidth]{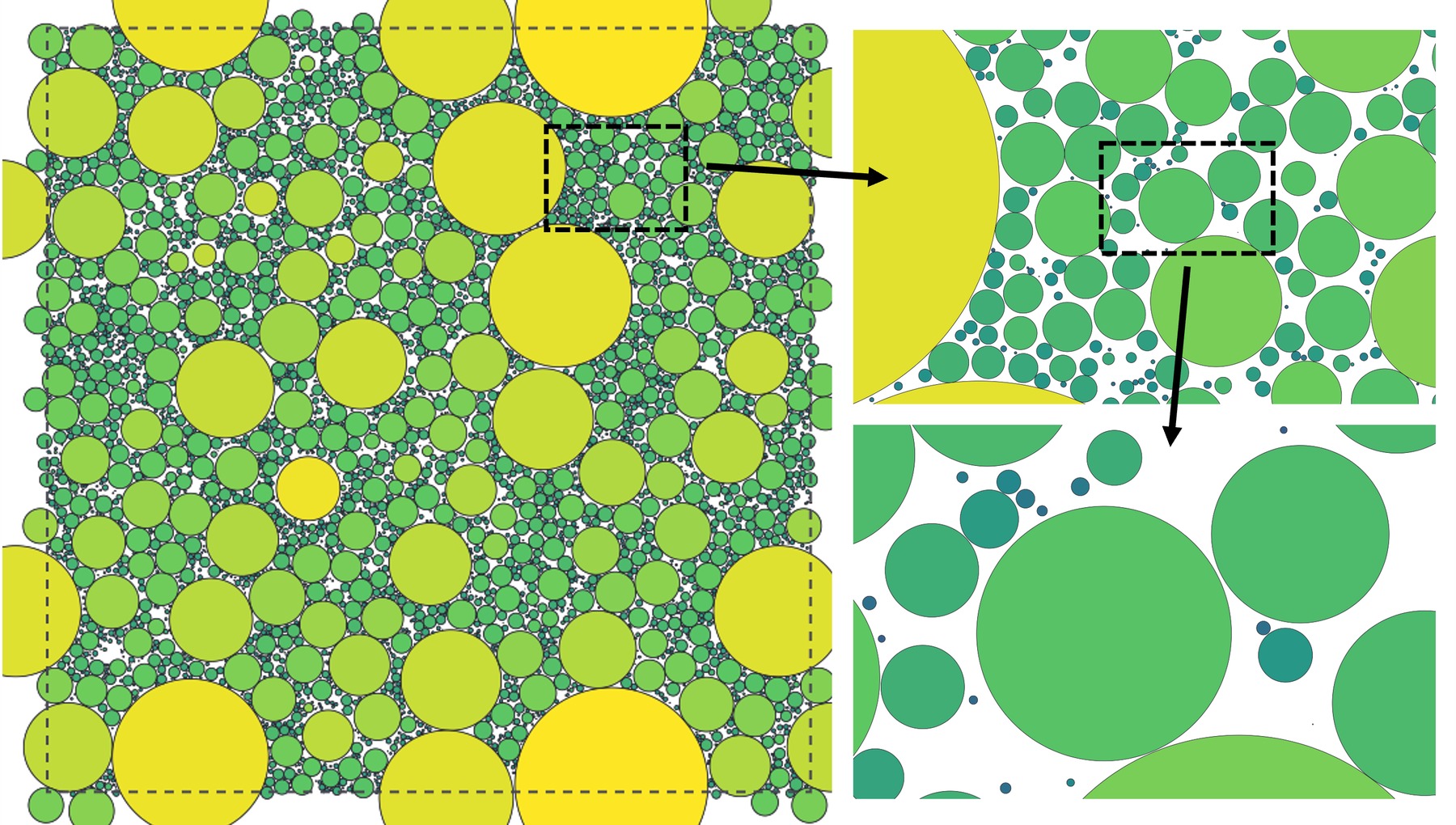}
	\caption{Planar slice through a truncated-lognormal random close packing at successive magnifications ($\alpha=1.1$, $a=4.75$; $S\approx3\times10^{4}$, $N=1.34\times10^{6}$, $\phi\approx0.80$, $D_{\max}/L=0.3$), each cut sphere drawn as disk and colored by 3D diameter (log scale).}
	\label{fig:slice_broad}
\end{figure*}

The truncated lognormal draws diameters as
\begin{equation}
D = D_g\,e^{\alpha z},\qquad D_g=1,\qquad |z|\le a,
\label{eq:lognormal}
\end{equation}
sampled by the equal-probability quantiles as discussed in Sec.~\ref{sec:algo}, with size ratio
\begin{equation}
S \equiv D_{\max}/D_{\min} = e^{2 a\alpha}.
\label{eq:S}
\end{equation}
The width $\alpha$ sets the polydispersity and the truncation $a$ controls how much of the heavy tail is retained, with $a\to\infty$ recovering the untruncated lognormal. We sweep $\alpha\in[0.1,1.3]$ and $a\in[3.5,5.25]$ in three dimensions. For each case the particle number $N$ is chosen such that the largest particle spans a fixed fraction of the container edge, $D_{\max}/L\in\{0.1,0.2,0.3,0.4\}$ depending on the run, with the broadest distributions reaching $N\approx3.1\times10^{6}$ spheres, and $L$ is the container width. All cases with fewer than one million particles are run three times, each from a different random seed, over which $\phi$ varies by only $\approx\!5\times10^{-4}$. To illustrate the resulting microstructure, Fig.~\ref{fig:slice_broad} shows a slice through one such packing. Since $P(D)$ is continuous, there is no discrete coarse scaffold with fines filling isolated voids; instead, the sizes grade smoothly from largest to smallest, with a few small pockets remaining locally underfilled.

\begin{figure}[b]\centering
	\includegraphics[width=\linewidth]{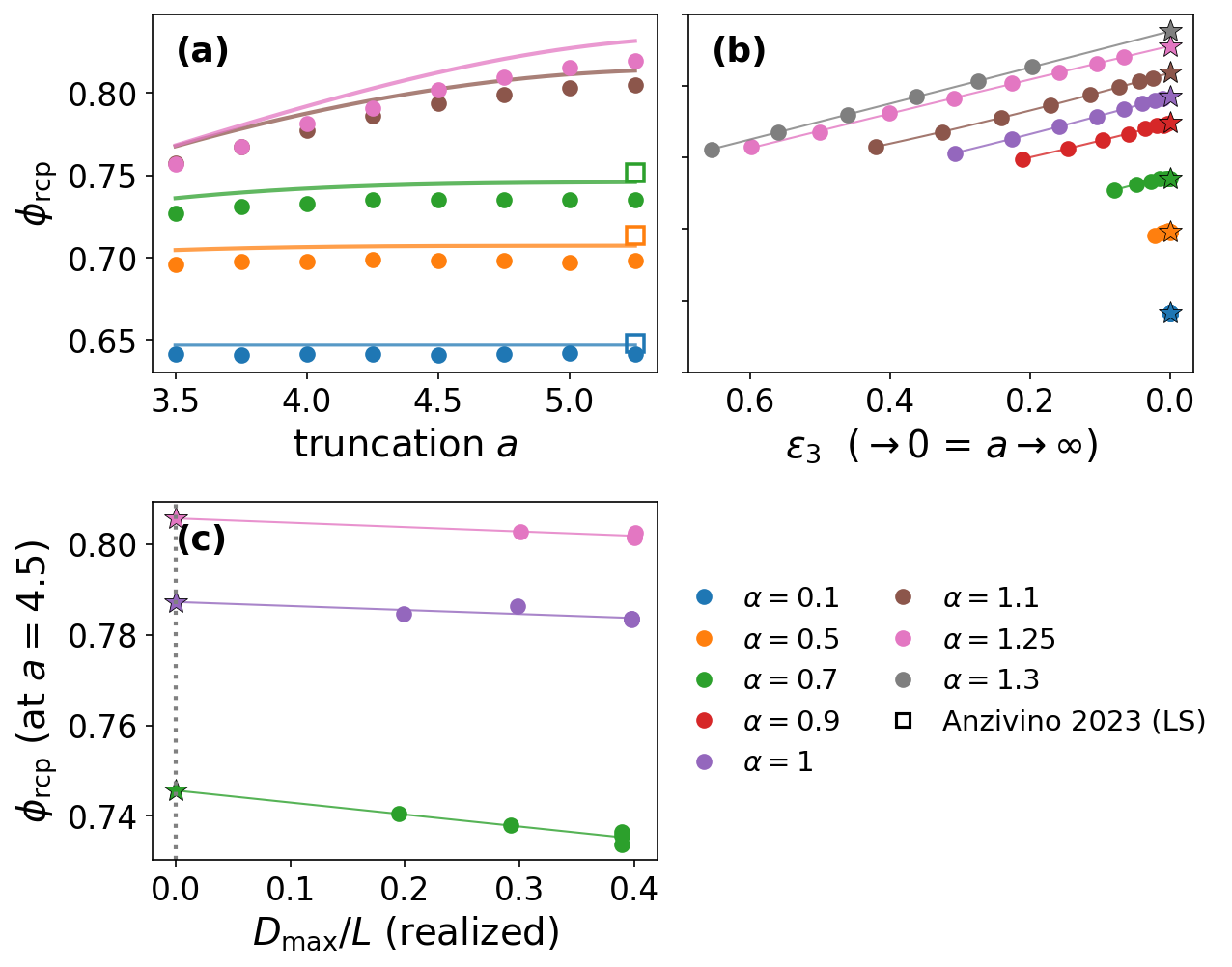}
	\caption{Lognormal sweeps, colored by width $\alpha$ (legend, lower right). \emph{(a)}~raw $\phi$ vs truncation $a$ (symbols; error bars span seeds), Farr--Groot for the same $(\alpha,a)$ (curves), and the Anzivino \emph{et al.}~\cite{anzivino_2023} hard-sphere simulations (open squares). \emph{(b)}~truncation extrapolation: $\phi$ vs the missing third-moment fraction $\epsilon_3$, with quadratic fits to the untruncated intercepts $\phi_{a\to\infty}(\alpha)$ (stars). \emph{(c)}~finite-size extrapolation: $\phi$ at $a=4.5$ vs $D_{\max}/L$, linear fits to $D_{\max}/L\to0$.}
	\label{fig:lnsweeps}
\end{figure}

Figure~\ref{fig:lnsweeps}(a) shows the raw $\phi$ against the truncation $a$ across widths, where the packing fraction rises with $a$. Reducing the truncation admits more of both the coarse and fine tails. The coarse particles carry most of the volume, and the fines fill the interstices. Additionally, increasing $\alpha$ results in a steeper rise in $\phi$, since $\alpha$ broadens the distribution and raises the polydispersity. The curves in the plot are the parameter-free prediction of Farr and Groot~\cite{farr_groot_2009,farr_2013} (Appendix~\ref{sec:fgtheory}), distinctive in returning $\phi$ for any $P(D)$ with nothing fit to our data, a purely geometric construction rather than a calibrated model. The data track it closely: at every $(\alpha,a)$ the prediction sits a nearly constant offset \emph{above} the measured points, where the simulation reproduces the predicted shape up to a small, systematic gap. As our generated packings are finite in size and the distributions are truncated, we anticipate some reduction in $\phi$ due to these effects, which we quantify next. We therefore read Farr--Groot as a yardstick for the \emph{shape} of $\phi$, not a target to match. As an independent check, the open squares in the plot are the Lubachevsky--Stillinger hard-sphere simulations of Anzivino \emph{et al.}~\cite{anzivino_2023}, which our packings track a little below. 

We quantify and remove each effect, finite size and truncation, beginning with the finite-size depression. To measure the finite-size effect contribution, we plot in Fig.~\ref{fig:lnsweeps}c $\phi$ as a function of $D_{\max}/L$ and fit a line to extrapolate $\phi$ to $D_{\max}/L\to0$, giving $\phi$ for an infinite container. Next we quantify the effects of truncating the lognormal distribution to a finite size ratio. We start by noting that a sphere's volume scales as $D^{3}$, so $\phi$ is carried by the large-diameter tail. Cutting it at finite $a$ removes part of that volume-bearing population; the fraction of the third moment $\langle D^{3}\rangle$ removed,
\begin{equation}
  \epsilon_3 = 1-\frac{\Phi(a-3\alpha)-\Phi(-a-3\alpha)}{\Phi(a)-\Phi(-a)},
  \label{eq:eps3}
\end{equation}
with $\Phi$ the standard normal cumulative distribution, measures this and vanishes as $a\to\infty$. A quadratic fit of $\phi$ against $\epsilon_3$ (Fig.~\ref{fig:lnsweeps}b) extrapolates to the untruncated value $\phi_{a\to\infty}(\alpha)$; applied to the Farr--Groot values themselves it recovers their known $a\to\infty$ limit to within $\approx\!0.006$. Only $\alpha=1.30$ is materially uncertain, as even the smallest computationally accessible truncation leaves $\epsilon_3\gtrsim0.2$, and the same check shows its intercept is, if anything, under- rather than over-estimated.

\begin{figure}[b]\centering
	\includegraphics[width=\linewidth]{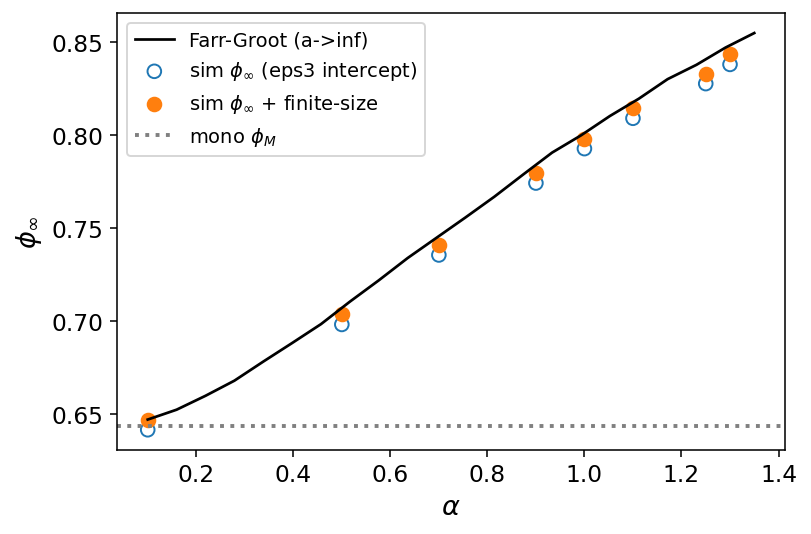}
	\caption{Line: Farr--Groot for the untruncated lognormal, calibrated only by $\phi_M=0.6435$. Open circles: $\phi_{a\to\infty}(\alpha)$ from the $\epsilon_3\to0$ extrapolation; filled circles: the doubly-extrapolated $\phi_\infty(\alpha)$ after the finite-size correction.}
	\label{fig:phiinf}
\end{figure}

Combining the two extrapolations yields a doubly-\emph{extrapolated} infinite-system estimate $\phi_\infty(\alpha)$, plotted against the parameter-free Farr--Groot prediction in Fig.~\ref{fig:phiinf}. We stress that $\phi_\infty$ is inferred from the finite-box data by these two fits, not measured directly. These corrected data reproduce the \emph{shape} of the prediction across $0.1\le\alpha\le1.3$ and lie consistently about $5\times10^{-3}$ below it. This offset is comparable to the $\approx0.006$ accuracy of that truncation extrapolation, so we do not read the residual as a firm discrepancy.

\subsection{Power law}\label{sec:sat}

The truncated power law is defined as $P(D)\propto D^{p}$ on $D\in[1,S]$ with size ratio $S\equiv D_{\max}/D_{\min}$ and exponent $p$ that sets the shape. Raising $S$ for a fixed $p$ has the effect of admitting ever larger particles; lowering $p$ shifts more of the sample's volume onto the smaller ones. We sweep $p$ from $-5$ to $-2$, a range that brackets the densest ``optimal grading'' near the Andreasen value $p\approx-3.63$~\cite{andreasen_andersen_1930,madani_rostami_2017,oquendo_estrada_2022}, and $S$ from $5$ to $200$ in three dimensions. As with the truncated lognormal, the particle number is determined such that $D_{\max}/L$ is $0.1$, $0.2$, or $0.4$. As in the lognormal test, samples of fewer than 1 million particles are generated 3 times from different initial seeds.

\begin{figure}[t]\centering
	\includegraphics[width=\linewidth]{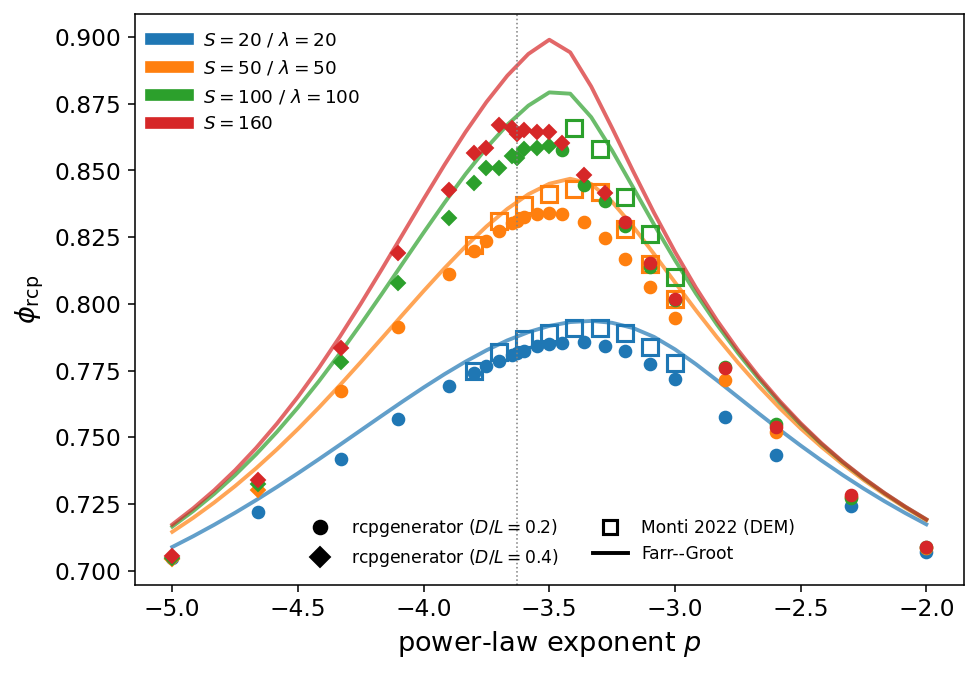}
	\caption{Measured $\phi(p)$ at $S=20$, $50$, $100$ and $160$ (filled: circles $D_{\max}/L=0.2$, diamonds $0.4$), the parameter-free Farr--Groot prediction for the same power law at the same $S$ (curves; no fit), and the DEM simulations of Monti \emph{et al.}~\cite{monti_2022} at the matched $\lambda=20$, $50$, $100$ (open squares; their data stop at $\lambda=100$, so $S=160$ carries only ours and the prediction). Dotted: the Andreasen exponent $p\approx-3.63$.}
	\label{fig:pwS}
\end{figure}

Figure~\ref{fig:pwS} shows the measured $\phi(p)$ at four size ratios, $S=20$, $50$, $100$ and $160$; the solid lines are the parameter-free Farr--Groot prediction for the same power law at the same $S$, with no fit. Both trace an asymmetric hump. The peaks in the data and theory are well aligned, and sit slightly to the right (larger $p$) of the Andreasen value. In both, the peak converges toward the Andreasen value with increasing $S$. Additionally, raising $S$ lifts $\phi$ across all $p$ as progressively larger grains are admitted. Inspecting the effects of finite size, we find it does not influence the shape of these curves or adjust the packing fraction by more than $\lesssim0.005$. The data do show a systematic offset, where our packing fractions lie below Farr--Groot by $\approx\!0.01$. This offset increases to $\approx\!0.03$ near the peak at $S=160$, where $\phi\simeq0.87$, indicating an apparent plateau in $\phi$ with $S$ that diverges more from theory near $p=-3.63$. 

As another comparison to our results, the open squares are the DEM simulations of Monti \emph{et al.}~\cite{monti_2022} at matched size ratio ($S = 20$, $50$ and $100$) and common $D_{\max}/L = 0.2$. The data show that our packing fractions consistently track slightly below the MD results, by $\approx\!0.005$--$0.015$, which also sit below Farr--Groot. That offset behaves as the offset to theory does, smallest at the steepest exponents, largest near the optimum, and growing with size ratio. Between all three methods---Adam quench, MD, Farr--Groot---the $\phi(p)$ curves are characteristically the same, separated by systematic offsets of order $0.01$.

\begin{figure}[b]\centering
	\includegraphics[width=\linewidth]{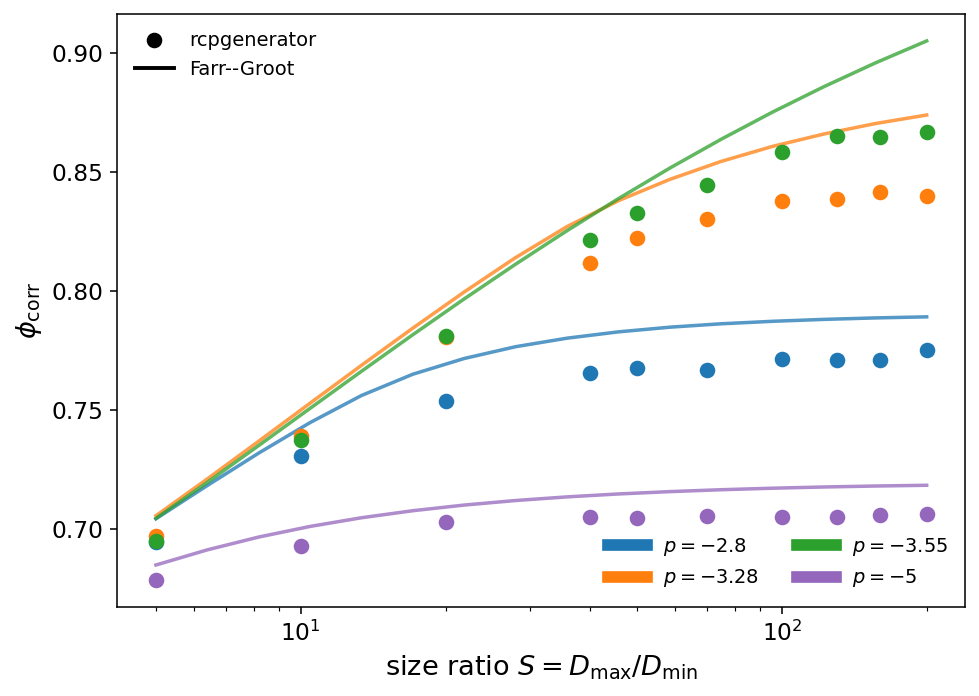}
	\caption{Measured $\phi(S)$ for $p=-2.8,-3.28,-3.55,-5$ (symbols) and Farr--Groot for the same exponent (curves; no fit).}
	\label{fig:pwSsat}
\end{figure}

Farr--Groot predicts that far from the peak $\phi$ saturates with $S$ at moderate size ratios, of order $100$ or less, while nearer the peak saturation is pushed out to larger and larger $S$. That is not what we find. Figure~\ref{fig:pwSsat} plots $\phi(S)$ at four exponents, two far from the optimum ($p=-5$ and $-2.8$) and two near it ($p=-3.28$ and $-3.55$), against Farr--Groot for the same exponent. Far from the optimum the two agree on where the rise ends: both flatten by $S\approx100$, and the offset simply holds, $\approx\!0.011$ at $p=-5$ and $\approx\!0.018$ at $p=-2.8$. Near the optimum they part. Our $\phi$ flattens by $S\approx160$ and gains nothing after, staying level to within $\pm0.005$ out to $S=200$, while the prediction climbs on, widening the offset from $\approx\!0.01$ at $S=20$ to $\approx\!0.04$ by $S=200$. We do not assert either to be correct: our data extend only as far as the available compute permits, and the prediction is untested at these size ratios. There is, however, reason to suspect $\phi$ should continue to climb as accommodating voids are present in the packings.

\begin{figure*}[t]\centering
	\begin{overpic}[width=0.8\textwidth]{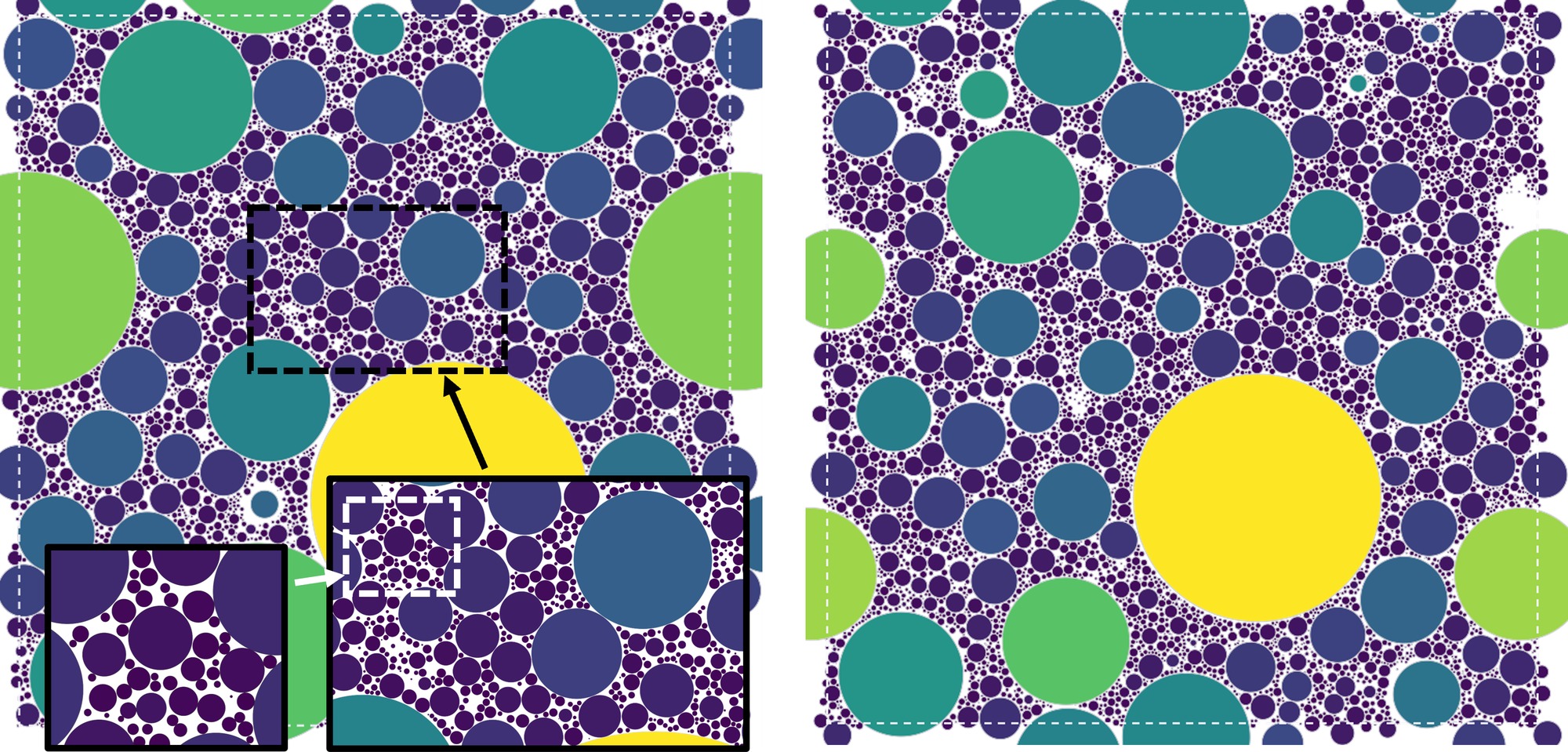}
		\put(47.6,44.6){\llap{\setlength{\fboxsep}{2pt}\colorbox{white}{\footnotesize (a)\ \ $z=0.32$}}}
		\put(98.6,44.6){\llap{\setlength{\fboxsep}{2pt}\colorbox{white}{\footnotesize (b)\ \ $z=0.42$}}}
	\end{overpic}
	\caption{\emph{(a)} and \emph{(b)} are two-dimensional slices through the same three-dimensional packing with a power-law size distribution at different depths, $p=-3.63$ and $S=100$ ($N\approx5.6\times10^{5}$, $\phi\approx0.86$), each cut sphere colored by 3D diameter (linear scale). \emph{(a)} is at depth $z=0.32$ and \emph{(b)} at $z=0.42$; the boxes in \emph{(a)} are successive zooms.}
	\label{fig:slice_optimum}
\end{figure*}

The power law near the Andreasen exponent is the classical \emph{optimal grading}: a self-similar distribution in which each size class fills the voids left by the next-larger one, a space-filling cascade that approaches an Apollonian filling as $S\to\infty$~\cite{furnas_1931,brouwers_2006,brouwers_2011,borkovec_paris_peikert_1994,anishchik_medvedev_1995,varrato_foffi_2011}. Each decade of $S$ should add another generation of fines, and the parameter-free prediction duly climbs well past $0.9$, toward $\phi\approx0.98$ as $S\to\infty$. However, the densest single packing we obtain is $\phi\approx0.87$, which has leveled off at $p\approx-3.6$, $S=160$ ($N\approx1.9\times10^{6}$). This grading has been identified numerically at moderate size ratios, including the frictionless case relevant here~\cite{oquendo_estrada_2020,oquendo_estrada_2022}. The present method extends the observation to $S=200$, where the plateau becomes visible.
We do not have an explanation for the plateau we observe, but inspection of the packing structure hints at one. Figure~\ref{fig:slice_optimum} shows two 2D slices through a 3D packing generated at $p=-3.63$ and $S=100$. Both are densely filled across every size class, as the grading intends. We also see, however, regions of open pore space between the coarse grains, leaving room for further increase in packing fraction if filled. These voids are not a defect: the configuration simply no longer rearranges under this protocol, so smaller particles do not fill the remaining voids. This is not to suggest that the Adam optimizer inhibits such denser packings, but simply that these are the results produced by the protocol used here. There may exist other formulations and implementations that retain the size-ratio stability and speed of this algorithm while enabling further densification, as suggested by swap-based approaches~\cite{ninarello_berthier_coslovich_2017,brito_lerner_wyart_2018}.

\subsection{Weibull}\label{sec:weibull}

The Weibull (Rosin--Rammler) law provides a contrasting third form,
\begin{equation}
P(D)\propto\left(\frac{D}{\lambda}\right)^{k-1}\exp\!\left[-\left(\frac{D}{\lambda}\right)^{k}\right],\qquad D\in[1/\sqrt{S},\,\sqrt{S}],
\label{eq:weibull}
\end{equation}
where the characteristic (Rosin--Rammler) scale $\lambda$ is the largest diameter, $\lambda=\sqrt{S}$. It is light-tailed and finite-variance, and its polydispersity is set by the modulus $k$ (small $k$ broad, large $k$ narrow) and \emph{not} by the size ratio. The form is empirical in origin, obtained from the analysis of ground coal~\cite{rosin_rammler_1933}, and it remains widely used to describe comminuted materials~\cite{alderliesten_2013}. We measure $\phi$ as a function of the modulus $k$ at fixed $S=300$, repeating each $k$ over a ladder of largest-particle box fractions, $D_{\max}/L\approx0.095$ down to $\approx0.019$ (the tightest boxes packing $N\approx3.1\times10^{6}$ spheres). We find $S$ inert: widening the support $[1/\sqrt{S},\sqrt{S}]$ adds negligible probability mass and produces no measurable change in $\phi$. From the sweeps in $D_{\max}/L$ we extrapolate $\phi$ to $D_{\max}/L\to0$.

\begin{figure}[t]\centering
  \includegraphics[width=\linewidth]{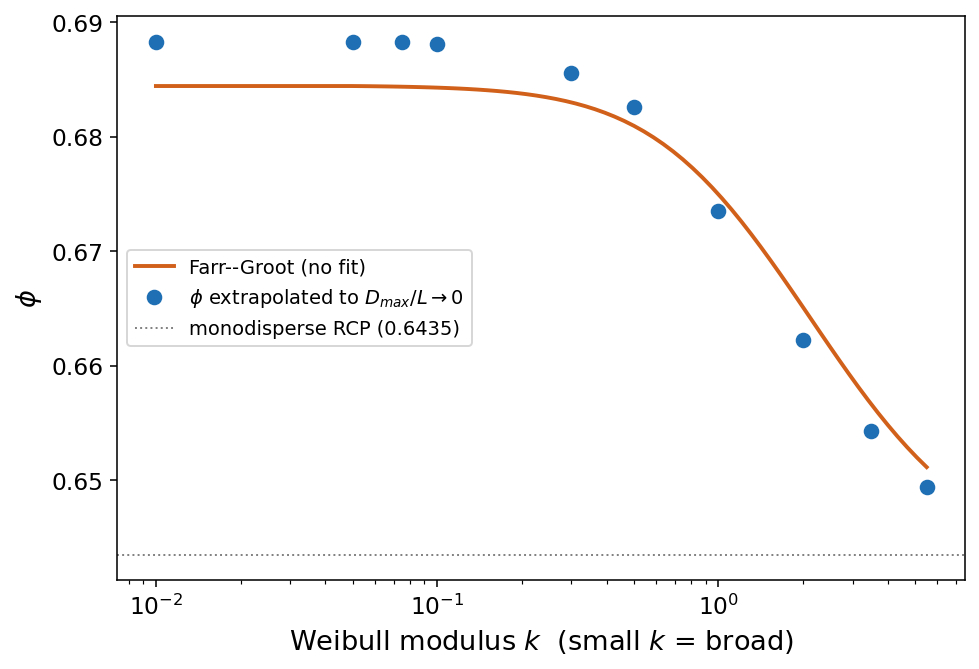}
  \caption{Infinite-box estimate of $\phi(k)$ at $S=300$ (symbols) versus the parameter-free Farr--Groot prediction for the same Rosin--Rammler law (curve; no fit). Dotted: the monodisperse value $0.6435$.}
  \label{fig:wbK}
\end{figure}

Figure~\ref{fig:wbK} shows the extrapolated $\phi(k)$ from the simulations, with the parameter-free Farr--Groot prediction as the solid curve. At large $k$ the distribution is narrow and $\phi$ sits just above the monodisperse value, reaching $0.650$ at $k=5.5$. Lowering $k$ broadens the distribution, with $\phi$ increasing monotonically, turning over at a knee near $k\approx0.3$ and saturating at $\phi\approx0.688$ below $k\approx0.1$. Farr--Groot predicts the same shape and the same knee, but differs in magnitude, and the difference \emph{changes sign} with width: the broad distributions sit slightly above the prediction (by $\approx0.004$), the narrow ones marginally below it (by $\approx0.002$--$0.003$). The Weibull thus \emph{brackets} the Farr--Groot value rather than tracking uniformly below it, unlike the lognormal (flat below) and the power law (a growing, one-sided deficit). 

\section{Comparison of multimodal $P(D)$ with experiment}\label{sec:experiment}

The distribution families tested so far are all \emph{continuous}. At large size ratio, however, a \emph{discontinuous} multimodal distribution can pack just as densely and, given the right ratio of species, even denser. Conceptually a multimodal packing is simple: it holds $n$ classes of particles, each of a characteristic size. The operating principle is that the classes form a discrete hierarchy of well-separated sizes, each filling the voids left by the next-larger one~\cite{furnas_1931,mcgeary_1961,brouwers_2006}: for $n$ such classes the void-filling cascade has been shown to give
\begin{equation}
\phi_n = 1-(1-\phi_1)^{n},
\label{eq:cascade}
\end{equation}
with $\phi_1\approx0.6435$ the monodisperse value, so $\phi_n\to1$ as $n\to\infty$ (Table~\ref{tab:cascade}). This is the route to dense-packed materials across engineering, for instance, graded concrete aggregate, ceramic and powder-metallurgy feedstocks, and additive-manufacturing powders~\cite{sobolev_2010,bai_2017}. The densest continuous packing we obtained, $\phi\approx0.87$, sits essentially at the $n=2$ rung, and a discrete ternary can in principle reach $\phi\approx0.95$ at $n=3$ with tailored preparation. Unlike the continuous families, this regime has a classic experiment to compare against: the vibrated multimodal packings of McGeary~\cite{mcgeary_1961}.

\begin{figure}[t]\centering
	\begin{overpic}[width=0.82\columnwidth]{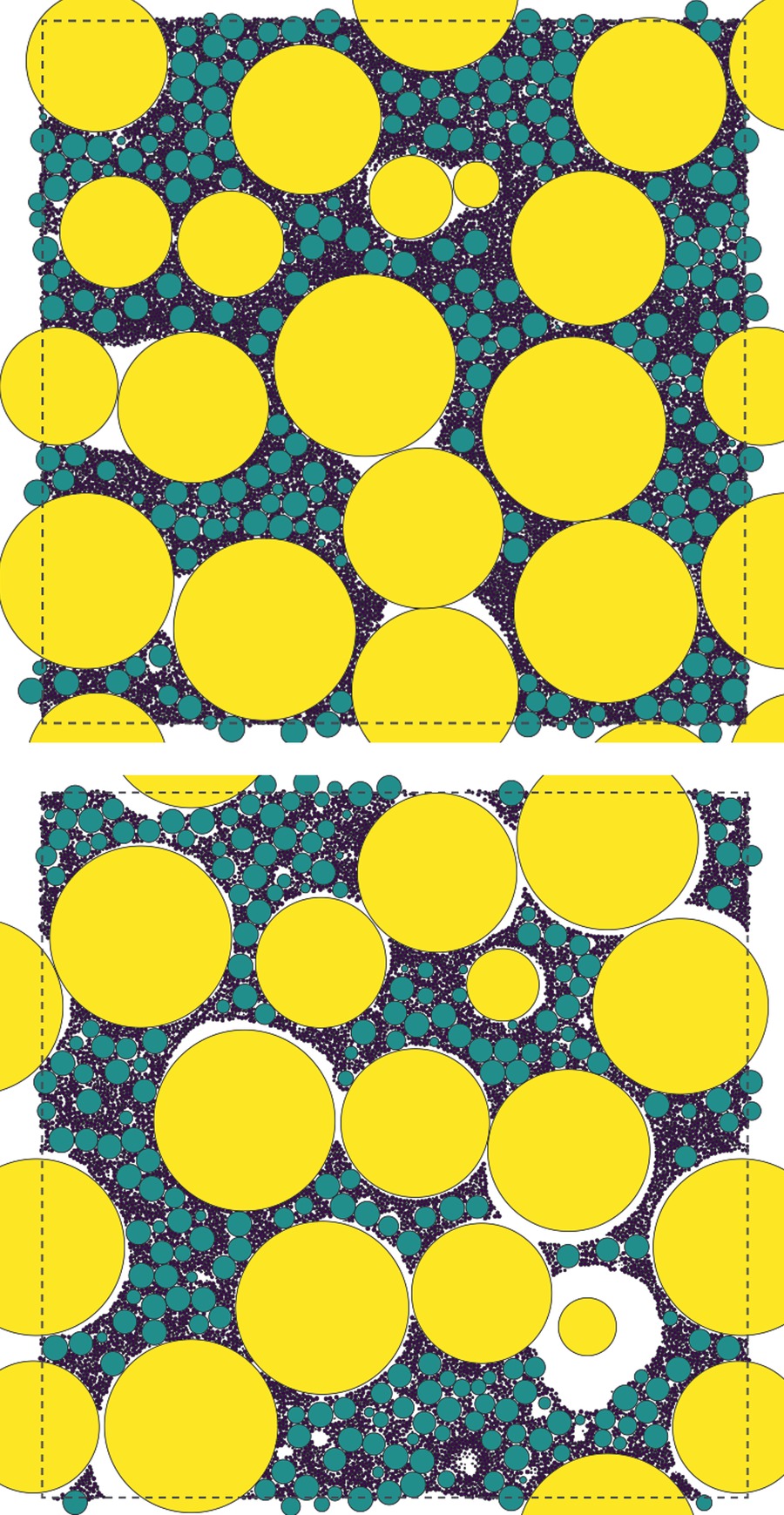}%
		\put(2.5,94){\colorbox{white}{(a)}}%
		\put(2.5,43){\colorbox{white}{(b)}}%
	\end{overpic}
	\caption{Planar slices through ternary packings of the \emph{same} grading, colored by size class. \emph{(a)} the sequential cascade ($\phi\approx0.88$); \emph{(b)} the same grading in a one-shot quench ($\phi\approx0.82$).}
	\label{fig:slice_cascade}
\end{figure}

In the experiments of McGeary, the sample is prepared via \emph{sequential} construction, built class by class. McGeary poured the largest spheres and vibrated them to a jammed bed; added the next-smaller class and vibrated it down into the interstitial pores until it too jammed; and repeated down the hierarchy. These ternary packings were constructed with diameter ratios $77{:}7{:}1$ (volume fractions $67{:}23{:}10$), and reached $\phi\approx0.90$ once prepared. The mechanism is staged: each finer class settles into the pore network the coarser one leaves behind, so the hierarchical filling proceeds one class at a time rather than all at once.

We reproduce the spirit of that construction, not McGeary's exact recipe. Starting from the coarsest class packed to its own dense state, we seat the next-finer class into the interstitial voids and inflate it to contact with the coarser class held fixed, then release everything to relax together (the numerical analogue of vibration) and repeat down the hierarchy. Our uniform $\rho\approx7$ and modest coarse counts differ from McGeary's larger, unequal ratios, so the comparison is by class count, not recipe.

Run this way the solver climbs the cascade: for a three-class grading at $\rho\approx7$ the packing fraction rises through $\phi_1=0.624\to\phi_2=0.800\to\phi_3=0.879$ (total $N$ up to $1.4\times10^{6}$, overlap-free), tracking McGeary rung by rung while remaining ${\sim}0.02$--$0.03$ below and approaching McGeary's $\phi\approx0.90$ ternary (Table~\ref{tab:cascade}). At these size ratios and species proportions, keeping the total particle count to near a million limits the coarsest class to 30, 45, or 60 particles, such that the packing density of the coarse single-class particles is $\phi_1=0.624$, carrying a finite-size depression below the bulk monodisperse $0.6435$. 

If instead the packings are prepared via a single pour (single shot), the same grading returns only $\phi\approx0.82$, below even the continuous optimum, because the coarse class arrests as sparse inclusions before the fines arrive, and the filling never proceeds. Figure~\ref{fig:slice_cascade} shows the contrast; staged, each class fills the voids of the one above, and one-shot, the coarse grains strand as sparse inclusions with large pores left open, where the configuration formed earlier in the protocol blocks the rearrangements that would let the fines reach them.

Even the cascade leaves the filling incomplete. Where the one-shot packing strands large pores empty of fines, the cascade leaves far fewer, but Figure~\ref{fig:slice_cascade} still shows small voids that no finer particle has reached, the fines settled at random rather than being steered into every pore as vibration would. One could go further in the same spirit, iterating the solver to seat additional particles into those residual voids and lifting the $n=2$ and $n=3$ fractions toward McGeary's $\approx0.90$. Optimizing that fill to match the experiment rung for rung was beyond our scope. The comparison already shows the solver reproducing the staged-construction mechanism and tracking the measured densities to within ${\sim}0.02$.

\begin{table}[t]\centering
  \begin{tabular}{lccc}
  \toprule
  size classes & ideal $\phi_n$ & McGeary $\phi_{\exp}$~\cite{mcgeary_1961} & $\phi$ (this work) \\
  \midrule
  1 (one size) & 0.644 & 0.625 & 0.624 \\
  2 (binary)   & 0.873 & 0.825 & 0.800 \\
  3 (ternary)  & 0.955 & 0.900 & 0.879 \\
  \midrule
  3 (ternary), one-shot & --- & ---$^{\dagger}$ & 0.82 \\
  \bottomrule
  \end{tabular}
  \caption{Packing fraction by number of well-separated size classes. \emph{Ideal $\phi_n$}: the void-filling cascade [Eq.~\eqref{eq:cascade}] with $\phi_1=0.6435$ (infinite class separation). \emph{McGeary}: vibrated metal-shot packings~\cite{mcgeary_1961}. \emph{This work}: the sequential-cascade generator (3D, $\rho\approx7$, $N_{\mathrm{big}}=30$--$60$, $N$ up to $1.4\times10^{6}$). Final row: the same ternary in a single one-shot quench; $^{\dagger}$McGeary reported the blended one-shot only qualitatively, so gives no value.}
  \label{tab:cascade}
\end{table}

\section{Conclusion}\label{sec:conclusion}\label{sec:limitations}
We have presented \texttt{rcpgenerator}, an openly available implementation that generates dense, disordered, \emph{non-overlapping} random close packings for a user-supplied list of diameters, dimension, and boundary geometry. Its reach is the broadly polydisperse regime that stalls conventional protocols: size ratios up to $S\sim5\times10^{5}$ for the lognormal family and particle numbers up to $N\approx5.6\times10^{6}$ for the power law, each packing produced in minutes to hours on a multicore machine. Broad polydispersity is made practical in this implementation by a per-coordinate adaptive optimizer, here Adam. Each iteration requires one force evaluation and position update, so runtime is governed mainly by the iteration count and hence by the largest stable step. For methods that use a single global step size, the growing stiffness contrast with $S$ generally restricts the largest stable step and raises the iteration count, and with it the cost. Adam instead adapts each coordinate, effectively preconditioning the update without Hessian information or line searches. It therefore retains a large stable step for each particle while keeping each iteration at the cost of a conventional force evaluation and position update, yielding the modest runtimes in Table~\ref{tab:cap} even at the largest $S$ and $N$. Other adaptive or preconditioned optimizers may perform similarly or better.

We compared the resulting packing fractions of our implementation against numerical results where they exist and against the parameter-free Farr--Groot prediction across three continuous families: truncated lognormal, truncated power law, and Weibull. In every case $\phi$ reproduces the distribution-controlled shape but differs systematically in magnitude, of order $0.005$--$0.01$, with a few power-law cases reaching $\approx0.03$. The offset depends on the distribution family: consistently about $5\times10^{-3}$ below for the lognormal, a few-percent one-sided undershoot that grows with $S$ for the power law, and a sign-changing bracket for the Weibull. Applying the method to mimic McGeary's sequential packing protocol for ternary packings reproduces the staged multimodal densities, rising rung by rung but again slightly lower in magnitude.

What the method does \emph{not} establish bears stating plainly. Random close packing is protocol-dependent~\cite{torquato_truskett_debenedetti_2000,ohern_silbert_liu_nagel_2003,xu_blawzdziewicz_ohern_2005,torquato_stillinger_2010}. The $\phi$ reported here are those of one inflation protocol, and we make no claim to a protocol-independent or maximally jammed value. The configurations are made non-overlapping but are not strictly or collectively jammed. The quantitative comparison is three-dimensional and periodic. The comparison to Farr--Groot serves throughout as a parameter-free \emph{yardstick}, independent of our generator but itself an approximation, so departures are reported as results, not verdicts on the theory. Nor is the density we obtain a ceiling: the packings sit slightly below the molecular-dynamics references and, in cross section, leave clearly unfilled voids. Therefore additional routes to modify the algorithm with the use of Adam or similar per-step adaptive methods may yet reach denser packings with little additional compute cost.

Beyond the packing fraction, the generator returns the full configuration, the particle positions and scaled diameters, at size ratios and system sizes other methods do not reach and where experiments record only bulk quantities. We release the code together with the complete per-case census behind every figure (see the Data and Code Availability section).

\section*{Data and code availability}\label{sec:data}
The \texttt{rcpgenerator} code, the analysis and figure-generation scripts,
and the size-distribution census underlying every figure are openly available in the maintained repository~\cite{rcpgenerator}; a frozen, self-contained archival snapshot of the code, census, scripts, and this manuscript is deposited at Zenodo, DOI:~\href{https://doi.org/10.5281/zenodo.21435447}{10.5281/zenodo.21435447}. To keep the archive compact, the
census is included directly (the per-case record of distribution parameters and resulting packing
fractions, from which all figures are produced), whereas the
individual packing positions and diameters are not, as they were not stored.
These packings are regenerable from the provided generation scripts and recorded random seeds. A single-thread timing benchmark (\texttt{scaling\_benchmark.csv}) supporting the quoted $N$-scaling is included. One reproducibility caveat applies: because the relaxation runs in parallel
across threads, generation is not bit-for-bit deterministic. The figures are
therefore exactly reproducible from the supplied census data, while regenerating
the packings from seed reproduces the reported quantities to within the run-to-run
scatter characterized in the text (a few parts in $10^{4}$ in $\phi$).

\bibliography{references}

\appendix

\section{Implementation details}\label{app:impl}

The solver takes a prescribed set of $N$ positive relative diameters
$\{D_i^0\}$ and places the particles without overlap at low volume fraction
using a cell-list procedure~\cite{allen_tildesley_2017}. The relative
diameters remain fixed throughout the calculation, while a single global
scale factor $\kappa$ sets the working diameters, $D_i=\kappa D_i^0$.
The optimized variables are therefore the particle positions
$\{\mathbf{x}_i\}$ and the scalar $\kappa$.

These variables minimize
\begin{equation}
	E(\{\mathbf{x}_i\},\kappa)
	=
	\frac{\epsilon}{2}\sum_{i<j}
	\left(d_{ij}-r_{ij}\right)^2
	\Theta\!\left(d_{ij}-r_{ij}\right)
	-\mu\sum_i D_i ,
	\label{eq:energy}
\end{equation}
where $d_{ij}=(D_i+D_j)/2$, $r_{ij}$ is the center-to-center
separation, and $\Theta$ is the Heaviside function. The first term is the
finite-range quadratic overlap energy used in the soft-sphere inflation
protocol~\cite{clarke_wiley_1987,xu_blawzdziewicz_ohern_2005,
	desmond_weeks_2009}; the second favors growth of the common diameter scale.
Optimization over the positions relaxes particle overlaps, while optimization
over $\kappa$ balances their elastic cost against the growth term. The parameter $\mu$ is chosen rather than optimized (here varied through a schedule), and the configuration we seek is the dense endpoint of this protocol, the largest $\kappa$ reached as $\mu\to0$.

The joint minimization is performed with Adam~\cite{kingma_ba_2015}, applied
to both the positional coordinates and $\kappa$. The global learning rate is
reduced during the calculation, with larger updates during initial
densification and smaller updates during the final relaxation. Because this dense endpoint is approached as $\mu\to0$, the schedule begins at larger $\mu$ and anneals it downward through a short sequence of discrete rungs, then cycles $\mu$ to promote further rearrangement. This schedule was hand-tuned rather than systematically studied, and likely leaves room for improvement.
Each rung runs for a prescribed optimizer-iteration budget, with the principal
transition set by
\begin{equation}
	n_{\mathrm{trig}}
	=
	4000\,Y+X\sqrt{Nd}-2000 ,
	\label{eq:trigger}
\end{equation}
where $d$ is the spatial dimension. The default schedule uses
$(Y,X)=(2,9)$ and the longer schedule uses $(Y,X)=(3,24)$, with a minimum
iteration count imposed for small systems. The generation scripts archived
with the code record the schedule used for each calculation.

At the final rung, relaxation stops when the maximum fractional pair overlap,
$1-r_{ij}/d_{ij}$, is below $5\times10^{-4}$. The remaining
geometric overlaps are then removed by reducing $\kappa$ uniformly, without
changing the particle positions or relative diameters, until the last overlap
vanishes. The solid-volume fraction of the resulting dense, disordered,
non-overlapping configuration is the reported packing
fraction $\phi_{\rm rcp}(P(D),N)$, written as $\phi$ in the main text and
stored as \texttt{phi\_corr} in the released census. Implementing the full method, including an adaptive learning rate on Adam, has many finer details and hyperparameters; for brevity we do not enumerate them here, but provide the source code as the definitive implementation reference.

Candidate interacting pairs are obtained from a $k$-d-tree spatial
index~\cite{bentley_1975}. The neighbor structure is rebuilt according to an
accumulated-displacement criterion rather than at every optimizer step.
The force loop uses batched neighbor queries, cache-oriented data storage,
and vectorized overlap evaluation. Together, these implementation choices
reduce the measured wall time by approximately a factor of ten relative to
the original direct implementation. The archived source code and generation
scripts provide the complete presets and implementation details used for the
reported calculations.

\section{The Farr--Groot prediction}\label{sec:fgtheory}
Farr and Groot~\cite{farr_groot_2009,farr_2013} predict the random-close-packing
fraction of any polydisperse sphere assembly by mapping the three-dimensional packing
onto an exactly solvable one-dimensional problem. A random line through the packing
intersects solid and void, on average, in the same proportion as the volume fraction
$\phi$; projected onto that line the packing becomes a sequence of one-dimensional rods
whose packing fraction equals the three-dimensional $\phi$. Each sphere of diameter $D$
is struck with probability proportional to its cross-section $D^{2}$ and contributes a
chord of length $L=D\sqrt{U}$ ($U$ uniform on $(0,1)$), so $P(D)$ fixes the rod-length
distribution $P(L)$ exactly. With the lengths thus determined, the only remaining
freedom is how the rods are arranged, or the gaps between them: Farr and Groot close the
problem with a greedy one-dimensional packing, which fixes the gap distribution given
$P(L)$, and the total gap length sets the one-dimensional packing fraction and hence
$\phi$. A single proportionality constant, calibrated once on the monodisperse limit
($\phi_M=0.6435$), maps the result back to three dimensions, after which the
construction has \emph{no} free parameter: given $P(D)$, the predicted $\phi$ follows
with no further input. We reimplement this mapping directly, reproducing the
monodisperse value by construction and the published trend at low-to-moderate
polydispersity, where it is well
supported~\cite{farr_groot_2009,farr_2013}, and
refer the reader to the original papers for the full derivation.

\end{document}